\documentstyle[twocolumn,prl,aps, epsfig]{revtex}

\begin{document}
\title{
A Droplet State in an Interacting Two-Dimensional Electron System
}
\author{Junren Shi$^{1}$, Song He$^{2}$, and X. C. Xie$^{1}$}
\address{
1. Department of Physics, Oklahoma State University,
Stillwater, OK 74078
}
\address{
2. Hexaa Laboratory, Warren, NJ 07059
}

\address{\rm (Draft on \today )}
\address{\mbox{ }}
\address{\parbox{14cm}{\rm \mbox{ }\mbox{ }
It is well known that the dielectric constant of two-dimensional (2D)
electron system goes negative at low electron densities.
A consequence of the negative dielectric constant could be
the formation of the droplet state. The droplet state is a
two-phase coexistence region of high density liquid and low
density "gas". In this paper, we carry out energetic calculations to study
the stability of the droplet ground state.
The possible relevance of the droplet state
to recently observed 2D metal-insulator transition is also discussed.
}}
\address{\mbox{ }}
\address{\parbox{14cm}{\rm PACS numbers: 71.30.+h, 73.40.Hm}}
\maketitle


\vspace{-0.5cm}

    The recent
discovery of two-dimensional (2D) metal-insulator transition (MIT)
by Kravchenko {\it et al.}\cite{Krav} has challenged the scaling
theory of localization\cite{Gang,Lee} in which a 2D MIT is forbidden.
A noticeable character of the electron system in these experiments is that $r_s$, the parameter measures the strength of the Coulomb interaction, is fairly large. We suspect that the electron system may be unstable against phase separation at these large values of $r_s$. We demonstrated in our previous paper\cite{He} that this assumption alone is sufficient to provide a theoretical description that is consistent with all the known experiments. For a two-dimensional (2D) electron system, there believed to be two phases:
a high density Fermi gas phase and a low density insulating
Wigner crystal phase. The dielectric constant of the liquid phase becomes negative when $r_s\simeq 2$\cite{Isihara}, which indicates that the liquid phase is unstable. 
At lower densities, the Wigner crystal phase
appears around $r_{s} \simeq 37$ in the absence of disorder\cite{Ceperley}.
This critical value of $r_s$
appears to be reduced to around $r_{s} \simeq 10$   
with disorders\cite{Chui}. In the intermediate values of $r_s$, we believe that there is a {\em liquid} phase which we think is responsible for the observed MIT.
 
In this paper, we propose that a droplet state of the electron system resulted from the phase separation of the electrons into this new liquid phase and a low density "gas" phase. 
Here we call the low density phase "gas" purely for the reason that its density is low.
In fact, in the presence of impurities, the "gas" phase is disordered Wigner crystal. To investigate our proposal, we have studied the energetics of such a droplet state. We find that both electron-electron
interaction and potential fluctuations are crucial for the formation
of the droplet state.

  An obvious condition for the droplet state is that the electron gas is unstable. To investigate what possibilities of the instability leads to, we study a simple but physically motivated model. Let us consider electrons in the disc of radius $b$ with positive background. Imagine that the electron system is shrunk to a new radius
$a<b$ while the positive background remains intact. Clearly the charging energy due to the separation of the electrons from the positive background increases the energy of the system.  However
there can also be energy gained (decreasing total energy):
since for a uniform electron gas the ground state energy $E_{g}$ is at
its minimum when $r_{s} \simeq 2$\cite{Isihara}, 
for $r_{s}>2$ the system            
gains energy by shrinking the area occupied by electrons.
Furthermore, in the presence of disorder, 
electrons tend to occupy the valleys
of the disorder landscape. Thus, a slowly varying
disorder potential is in favor for the formation of the droplet state.
We calculate the energy changes when electron disc is shrunk from
$b$ to $a$ to determine whether a spontaneous shrinking can take place.
 
For electron-electron or electron and positive background interaction,
we use screened Coulomb potential.
For $Si$ MOSFETs, the image charge in the metal 
substrate induces the screening and the 
interaction in the momentum space can be written as\cite{meissner}
\[
V(k)=\frac{1}{\varepsilon }\frac{2\pi e^{2}}{k}\frac{1-e^{-2kD}}{1-Ke^{-2kD}},
\]
where $D$ is the thickness of the \( Si_{2}O \) insulating layer and
\( 
\varepsilon =\frac{1}{2}\left( \varepsilon _{1}+\varepsilon _{2}\right)  \),
\( K=(\varepsilon _{1}-\varepsilon _{2})/(\varepsilon _{1}+\varepsilon _{2}) 
\), with
\( \varepsilon _{1} \) and \( \varepsilon _{2} \) being 
the dielectric constants
of $ Si$ and \( Si_{2}O \), respectively.
For other systems, such as $GaAs/AlGaAs$, 
the screened interaction can be well 
represented by the following form:
\[
V(r)=\frac{e^{2}}{\varepsilon r}e^{-\lambda r},\]
and the corresponding moment space representation is,
\[
V(k)=\frac{2\pi e^{2}}{\varepsilon }\frac{1}{\sqrt{k^{2}+\lambda ^{2}}}.\]
Both forms of the interactions define an interaction range \( \xi  \). For
$Si$ MOSFETs, \( \xi =\frac{\varepsilon }{\varepsilon _{2}}D \), and
for the screened Coulomb potential, \( \xi =\frac{1}{2\lambda } \). Outside 
the range the interaction is strongly screened.

\begin{figure}[h]
{\centering
\epsfig{file=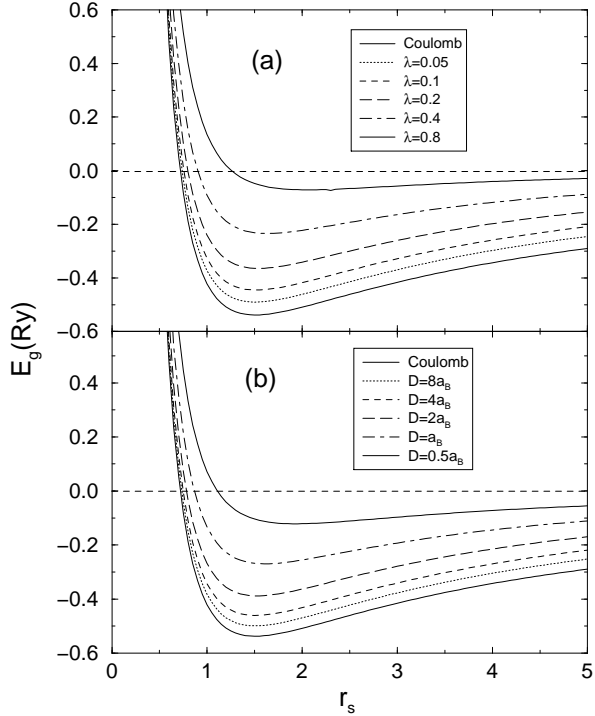, width=0.9\columnwidth}
\par}
\caption{\label{fig: 1}The ground state energy of electron system for
different screening
strength: (a) for screened Coulomb interaction, (b) for Si MOSFETs.}
\end{figure}

To calculate the ground state energy of the electron gas with the screened
Coulomb interaction, we use the variational correlated-basis-function (CBF) 
method\cite{cbf}. This method has been applied to the bare Coulomb interaction
and is proved to provide rather accurate result for the ground state energy
of 2D electron system\cite{cbf}. 
The accuracy of the ground state energy is found to within $10\%$ comparing
with the best available quantum Monte-Carlo results\cite{Ceperley} for
densities down to $r_{s}=20$.
In the CBF approach, there is a variational variable $\alpha$ for which
the ground state energy $E_{g}$ has to be minimized

\begin{eqnarray*}
E_{g}(r_{s},\alpha ) & = & \frac{A(\alpha )}{r_{s}^{2}}-U_{c}
\left( \frac{\sqrt{\alpha }}{r_{s}}\right) \, \, \, \, \, \, \, (Ry.)\\
A(\alpha ) & = & A_{0}^{B}(\alpha )+A_{01}^{F}(\alpha )+A_{02}^{F}(\alpha )
+A_{03}^{F}(\alpha )+\cdots \\
A_{0}^{B}(\alpha ) & = & \alpha ^{2}\left[ \sum _{n=0}^{\infty }\frac{\alpha 
^{n}}{(n+2)^{2}}+\frac{\pi ^{2}}{6}-\frac{5}{4}\right] \\
A_{01}^{F}(\alpha ) & = & 1\\
A_{02}^{F}(\alpha ) & = & -\frac{16}{\pi }\int _{0}^{1}\left[ 
2\arccos (y)-y(1-y^{2})^{1/2}\right]  \\
& & \times y^{3}e^{-2y^{2}/\alpha }dy\\
A_{03}^{F}(\alpha ) & = & -\frac{2}{\pi }\int _{0}^{1}dy_{1}
\int _{0}^{1}dy_{2}\int _{0}^{1}dy_{3}\int _{0}^{\pi }
d\theta y_{12}^{2}\\
& &
\times   \left\{ 1-\exp \left[ -y_{12}^{2}/2\alpha 
\right] \right\} \\
 &  & \times \exp \left[ -\left( y_{1}^{2}+y_{2}^{2}+2y_{3}^{2}\right)
 /2\alpha \right] \\
 &  & \times I_{0}\left[ \alpha ^{-1}
\left( y_{1}^{2}y_{3}^{2}+y_{2}^{2}y_{3}^{2}+2y_{1}y_{2}y_{3}^{2}
\cos \theta \right) ^{1/2}\right], 
\end{eqnarray*}
where \( 1Ry=\frac{e^{2}}{2\varepsilon a_{B}},\, 
\, a_{B}=\frac{\varepsilon \hbar ^{2}}{m^{*}e^{2}} \), and
$y_{12}=y_{1}-y_{2}$.
\( U_{c}\left( \frac{\sqrt{\alpha }}{r_{s}}\right)  \) is the cohesive energy
which depends on the special form of the interaction,
\begin{eqnarray*}
U_{c}\left( \frac{\sqrt{\alpha }}{r_{s}}\right)  & = & \frac{1}{2}\int \exp
 \left[ -\left( \frac{k}{\frac{2\sqrt{\alpha }}{r_{s}}}\right) ^{2}\right]
 V(k)\frac{d^{2}k}{(2\pi )^{2}}.
\end{eqnarray*}

Figure \ref{fig: 1} shows the calculated results for the ground state energy.
The screening
effect raises the ground state energy because the electron correlation is
suppressed. 

The shrinking of the electron disc will cause the 
redistribution
of the charge, which will raise extra electrostatic energy
because the positive background is fixed. 
The charge
distribution can be written as

\[
\rho (k)=2N\left( \frac{J_{1}(ka)}{ka}-\frac{J_{1}(kb)}{kb}\right), \]
where \( N \) is the total number of the electrons in the disk. 
The charging energy is
\begin{eqnarray*}
E _{c} & = & \frac{1}{2N}\int \frac{d^{2}k}{(2\pi )^{2}}V(k)|
\rho (k)|^{2}\\
& = & \frac{1}{\pi a_{B}^{2}r_{s}^{2}}\int _{0}^{\infty }\frac{dx}{x}V\left(
\frac{x}{b}\right) \left( \frac{J_{1}(\gamma x)}{\gamma }-J_{1}(x)\right) ^{2},
\end{eqnarray*}
where \( \gamma =\frac{a}{b} \) is the ratio of the radii
after and before shrinking electron disc. The 
charging energy shows distinct forms 
for different interaction ranges. When the 
interaction
range \( \xi  \) is much larger than the radius of the disk, 
$\xi\gg b$, 
the electron-electron
interaction can be roughly considered as the bare Coulomb interaction. In
this case, the 
electrostatic charging energy is
\[
E_{c} 
 \approx   \frac{4b}{r_{s0}^{2}}\left( 0.290545-\frac{1}{\pi }\ln |1-
\gamma |\right) (1-\gamma )^{2}+\cdots \, \, \, (Ry).
\]
In the other limit $ b\gg \xi$, the interaction is well screened and
the electrostatic energy has the form
\[
E_{c} 
 =  4\xi \left| \frac{1}{r_{s}^{2}}-\frac{1}{ r_{s}^{\prime 2}}
\right| \, \, \, Ry,
\]
where \( r_{s} \) and \( r'_{s} \) are the inverse density
parameters before and after the shrinking. 
The total energy difference can be written as
\[
\Delta E_{tot}=E_{c}(\gamma )+\Delta E_{g}.\]
For a small initial radius $b\ll \xi$, the energy gain $\Delta E_{g}$ 
dominates over the energy loss $E_{c}$,
thus, there is always finite shrinking.

\begin{figure}[h]
{\centering 
\epsfig{file=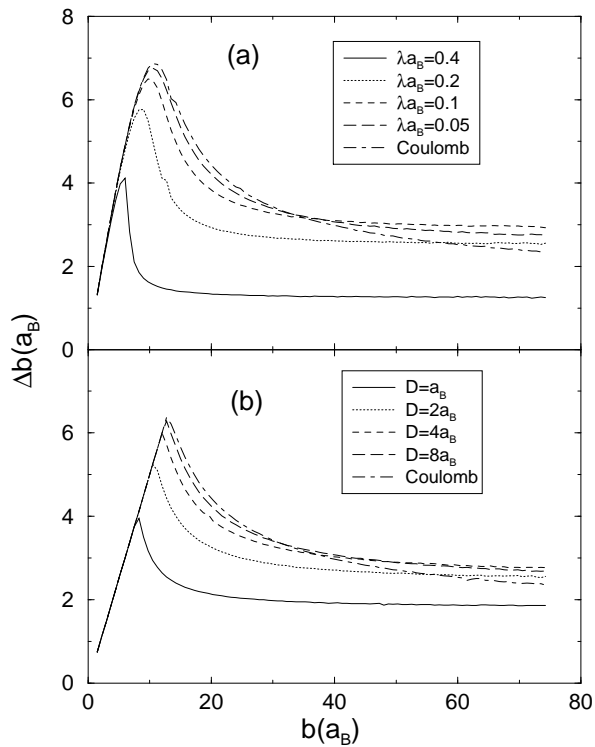, width=0.9\columnwidth}
\par}
\caption{\label{fig: 2}The shrinking distance $\Delta b= b-a$ 
versus disc radius \protect\( b\protect \) for different screening strength.
(a) for screened Coulomb potential; (b) for $Si$ MOSFETs. The initial 
density parameter \protect\( r_{s}=15a_{B}\protect \).}
\end{figure}

However, the above conclusion is not true in general as
demonstrated in Fig.\ref{fig: 2}.
Figure \ref{fig: 2} shows the shrinking distance $\Delta b= b-a$
versus initial radius $b$. 
$\Delta b$ approaches a constant for large $b$.
However, the shrinking shown here can
not be considered as a macroscopic shrinking 
because the typical $\Delta b$
is only about $2a_{B}$, 
which is far smaller than the average distance between the
electrons, which is $r_{s}=15a_{B}$ for this calculation. 
Similar behavior has also been
observed for other values of $r_{s}$.

Thus the intrinsic instability is not sufficient to
overcome the charging energy cost in order to form the electron droplet state.
The system is in a marginally stable situation. However, in real systems,
there are always disorders. 
The low frequency component of a disorder potential
forms the potential landscape, and electrons tend to occupy
the low potential valleys. We assume that around each local 
minimum, the disorder potential is isotropic and slowly varying.
We expand the disorder potential around the local minimum
up to the quadratic term. Therefore, we adopt the following 
simple model for disorder potential
\[
W(r)=V_{0}\frac{r^{2}}{b^{2}},\]
where $V_{0}$ is the potential depth from center to edge of the disc.
The energy gain by shrinking to a radius $ a$ can be evaluated as
\begin{eqnarray*}
\Delta E_{W} & = & \Delta \frac{1}{N}\int W(r)\rho (r)dS\\  
 & = & \frac{V_{0}}{2}\left( \gamma ^{2}-1\right). 
\end{eqnarray*}
The total energy difference in the limit $b\gg \xi$ can be written as

\[
\Delta E_{tot}=-\left( \frac{V_{0}}{2}-\frac{4\xi }{r_{s0}^{2}}\right)
 (1-\gamma ^{2})+\Delta E_{g}.\]
The effect of the electrostatic energy will be suppressed by the potential 
fluctuation.
Large value of $ V_{0}$ is in favor for the disc to shrink.

\begin{figure}[h]
{\centering
\epsfig{file=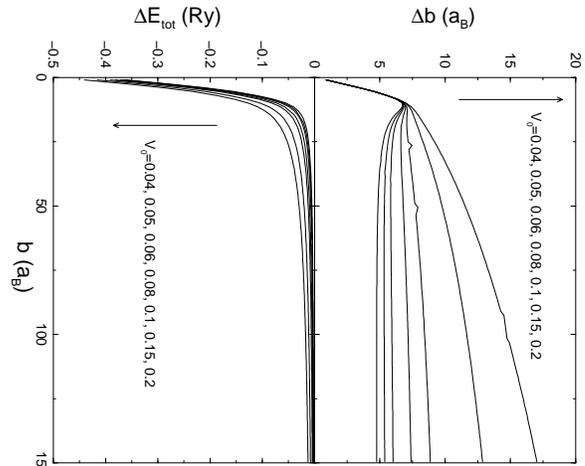, angle=270,width=0.9\columnwidth}
\par}
\caption{\label{fig: 3} (a) The shrinking distance $\Delta b= b-a$
versus radius \protect\( b\protect \) for $Si$ MOSFETs with
$D=10a_{B}$. Top curves are for larger disorder potential $V_{0}$'s.
(b) $\Delta E_{tot}$ versus radius \protect\( b\protect \).
Lower curves are for larger disorder potential $V_{0}$'s.
The initial electron density parameter \protect\( r_{s}=15a_{B}\protect \).}
\end{figure}                                     

 Fig.3(a) plots the $\Delta b=b-a$ as a function of the
initial radius $b$ for $Si$ MOSFETs with
$D=10a_{B}$ and $r_{s}=15a_{B}$. There exists a critical $V_{0}^c$ 
($\sim 0.08$) above which $\Delta b \propto b$. Thus, for large $b$ there is
a macroscopic shrinking for $V_{0}>V_{0}^c$. Similar result has also been
obtained for the screened Coulomb interaction with $1/\lambda$ places the
role of $D$. In Fig.3(b) we plot the energy change as a function of
the initial radius $b$. It is clear that larger value of $V_0$
gives rise to larger energy gain.
We have carried out the calculations for many values of
$r_{s}$ and the resulting phase diagrams are plotted in Figure 4.

\begin{figure}[h]
{\centering
\epsfig{ file=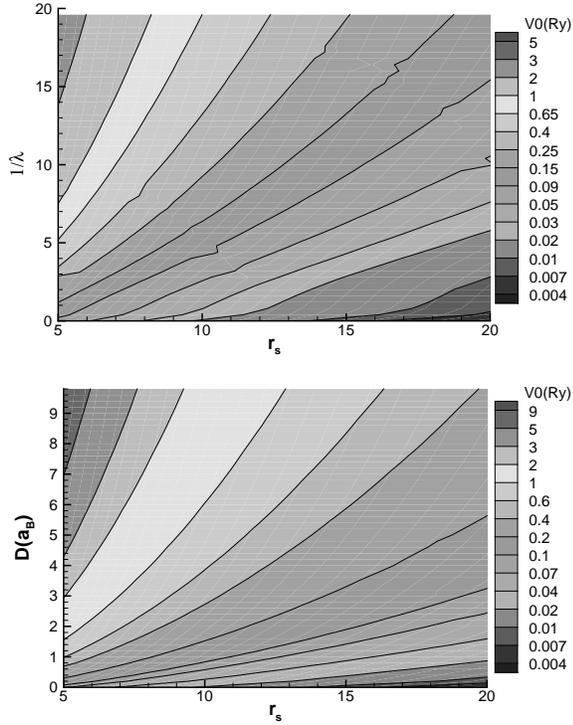, width=0.9\columnwidth}
\par
}
\caption{\label{fig: 4}The phase diagram in the $r_{s}-\xi$ plot.
$r_{s}$ is the initial density parameter of the electron disc.
$D$ or $1/\lambda$ represents
the interacting range $\xi$ of the electron-electron interaction.
Each value of $ V_{0}$
corresponds to a curve in the figure. On the right side of the curve,
the electron disc will have a macroscopic shrinking.
}
\end{figure}

Figure \ref{fig: 4} shows the phase diagram in the \( r_{s}-\xi  \) plot
for the case of $b\gg \xi$.
Each value of $ V_{0}$
corresponds to a curve in the figure. The curves for larger $V_{0}$
are above those for smaller $V_{0}$. On the right side of the curve
for a given $V_{0}$,
the electron disc will have a macroscopic shrinking, thus, an
electron droplet phase is stable.     
To form the electron droplet state, the screening of the electron-electron
interaction and the potential fluctuation are both
crucial. The smaller interaction
range between the electrons and the lower electron density, the
easier to form the droplet phase. 

Before summarizing, we would like to make several comments.
(i) In this paper, we only consider one electron disc, corresponding
to one drop in the droplet state. In real systems, electrons tend
to occupy valleys of potential fluctuations to give rise many
such drops. The size of each drop is determined by local potential
depth. (ii) We only consider zero temperature effect in this paper
such that the "gas" phase is empty. At a finite temperature,
The "gas" phase is occupied by lower density electrons. Thus, a finite
temperature enhances the possibility for the droplet state
since the density difference between the liquid and the "gas" is smaller
and the charging energy is less costly. However, in order to form
the droplet state, the temperature has to be below the cohesive
energy (the energy cost to remove an electron
from the liquid phase to the "gas" phase\cite{He}).
(iii) We believe that the recently observed 2D metal-insulator
transition might be the percolation transition of the liquid phase
in the droplet state\cite{He}. The percolation here is
semi-quantum in nature, different from the conventional classic
percolation\cite{Shi}. 
(iv) In order to have a percolation, the "gas" phase needs to have a
much smaller local conductivity than the liquid phase. This requires
that a typical length scale of the "gas" region is larger than
the localization length of the "gas" phase. The "gas" phase is
low in electron density which gives rise to a short localization length.
Thus, one may not need a large shrinking of electron drops to realize
a percolation transition.
We suspect that in experimental samples,
a typical drop size is of the order of $\mu m$. 

In summary, we have demonstrated that it is possible
to have a droplet phase for 2D electron systems at low
densities. Both electron-electron interaction and disorder
potential fluctuations are important for the formation of
the droplet phase.

The work is supported by DOE. 


\end{document}